# Designing Covalent Organic Framework-based Light-driven Microswimmers towards Intraocular Theranostic Applications

*Varun Sridhar[1,+], Erdost Yildiz[1,+], Andrés Rodríguez-Camargo,[2,3], Xianglong Lyu[1], Liang Yao[2], Paul Wrede[1], Amirreza Aghakhani[1], Mukrime Birgul Akolpoglu[1], Filip Podjaski[2,4,5,*], Bettina V. Lotsch[2,3,5,6,*], Metin Sitti[1,7,8,*]*

[1] Physical Intelligence Department, Max Planck Institute for Intelligent Systems, 70569 Stuttgart, Germany

[2] Nanochemistry Department, Max Planck Institute for Solid State Research, 70569 Stuttgart, Germany

[3] Department of Chemistry, University of Stuttgart, 70569 Stuttgart, Germany

[4] Department of Chemistry, Imperial College London, W12 0BZ London, United Kingdom

[5] Cluster of Excellence E-conversion, Lichtenbergstrasse 4, 85748 Garching, Germany

[6] Department of Chemistry, University of Munich (LMU), Munich, Germany

[7] Institute for Biomedical Engineering, ETH Zurich, 8092 Zurich, Switzerland

[8] School of Medicine and College of Engineering, Koç University, 34450 Istanbul, Turkey

[+] These authors contributed equally to this article.

[*] Correspondence to: sitti@is.mpg.de, f.podjaski@imperial.ac.uk, b.lotsch@fkf.mpg.de

**Abstract**

Even micromachines with tailored functionalities enable targeted therapeutic applications in biological environments, their controlled motion in biological media and drug delivery functions usually require sophisticated designs and complex propulsion apparatuses for practical applications. Covalent organic frameworks (COFs), new chemically versatile and nanoporous materials, offer microscale multi-purpose solutions, which are not explored in light-driven micromachines. We describe and compare two different types of COFs, uniformly spherical TABP-PDA-COF sub-micron particles and texturally highly nanoporous, irregular, micron-sized TpAzo-COF particles as light-driven microrobots. They can be used as highly efficient visible-light-driven drug carriers in aqueous ionic and cellular media, even in intraocular fluids. Their absorption ranging down to red light enables phototaxis even in deeper biological media and the organic nature of COFs enables their biocompatibility. The inherently porous structure with ~2.5 nm structural pores, and large surface areas allow for targeted and efficient drug loading even for insoluble drugs and peptides, which can be released on demand. Also, indocyanine green (ICG) dye loading in the pores enables photoacoustic imaging or optical coherence tomography and hyperthermia in *operando* conditions. The real-time visualization of the drug-loaded COF microswimmers enables new insights into the function of porous organic micromachines, which will be useful to solve various drug delivery problems.

**Introduction**

Microrobots are tiny machines that are tailored to be controlled externally to perform individual tasks. In order to achieve external control as well as multi-purpose functionality, micro/nanobots typically require a sophisticated and specifically adapted design enabling targeted control and applications. Their primary area of use is the large field of biomedicine.[1, 2, 3, 4, 5] The microrobots should not elicit any immune response and should be compatible with the cells to enable biomedical applications.[6, 7] Also, the propulsion method should be as noninvasive as possible, and non-toxic, excluding the use of dedicated toxic fuels.[8, 9] For an efficient microrobot function, motion control is the first requirement, which can become more and more challenging if the liquid they are propelled in contains species that hinder propulsion or external control. Wireless motion control requires external energy input and is typically realized by magnetic or acoustic actuation,[10] but can also be realized by ultraviolet (UV) or blue light, even in biological conditions, as evidenced very recently.[11] However, UV light, which is typically used for light-driven microswimmers[12, 13] is incompatible with biological tissues. Also, visible light control, usually reported with high-intensity blue light,[12] limits applications to transparent conditions since tissue penetration requires more red light or near-infrared light, which is a significant challenge to the field.

The critical tasks of mobile microrobots are cargo uptake and delivery, often linked to biopharmaceutical classes and properties of drugs, after actively navigating to a target diseased tissue region.[4, 7, 14, 15] Drug uptake and its controlled release were typically realized efficiently by encapsulation structures being a separate part of the microrobots; these were then opened in the desired conditions or where a release could be triggered otherwise. More recently, inherently porous structures, such as metal-organic frameworks[16, 17, 18] and porous carbon nitrides were used for such applications since their sizeable inner pore volume, reminiscent of a sponge, enables high and, even environmentally stable drug loading.[11] However, porous particle structures with many textural pores of different sizes as part of their inner surface area leave challenges for controlled loading and release from their volume.

As a last critical step to clinical applications, cell viability, the absence of foreign body reactions, and tissue biocompatibility are necessary conditions for microrobots to be used in biological contexts, which is not always easy to ensure, with all the desired functions being fulfilled at a time.[4, 7, 14] For this purpose, typically biocompatible metal coatings, such as gold, titanium, or polymers are employed, but also organic-based materials are up-and-coming and were used recently without coatings, such as carbon nitrides.[11, 17, 19] Compared to inorganic structures, organic materials not only offer potential biodegradability but also the high flexibility of chemical design of organic materials, especially in terms of surface functionalities and porosity, might enable more efficient and targeted biomedical applications, such as drug delivery or hyperthermia.[20] Even the most sophisticated designs with biocompatible metallic structures fail during actuation inside heterogenic biological fluids and specific-targeted drug delivery and imaging in live tissues.[21, 22, 23] Because of that, new materials and actuation methods should be investigated for the basic tasks for the clinical obstacles, such as intraocular motion and drug delivery.

In this work, we introduce covalent organic frameworks (COFs) as a tailorable active component to the field of micro- and nanomachines, or more precisely, light-driven microswimmers. These highly porous and crystalline materials can fulfill all the requirements listed above since their molecular structure, and also their morphology, can be designed and tuned bottom-up while enabling targeted properties.[20, 24, 25] Simultaneously, they can use visible light for photocatalytic reactions with their environment, which can also be used for active particle propulsion.[19, 26, 27] Depending on the propulsion mechanism and particle structure, the propulsion can be self-diffusiophoretic or self-electrophoretic, while allowing light-induced directional motion control via phototaxis.[28] The tailorable properties of the COF building blocks enable tuning of not only their absorption wavelength but also pore size and volume, surface polarity, and chemical affinity, which enable the loading of large and small molecule cargo based on the specific application requirements. Since such organic structures are non-magnetic per se, unless so-called Janus or hybrid (encapsulation) structures were to be employed, the use of light is a highly promising and convenient method not only to propel them but also to trigger functions within them and to image their behavior.[29] Thanks to their promising and ion-tolerant visible light-driven propulsion properties, we especially focused on their use in ophthalmological applications. In addition to their light-driven propulsion, their configurable particle sizes enable them to pass through the fibrillar mesh of vitreous humor (~500 nm pore size).[30] For this purpose, we selected therapeutic agents and imaging modalities accordingly.

Here, these possibilities are investigated and exemplified. We study and compare two very different modified COFs, namely TAPB-PDA-COF made from the condensation of 1,3,5-tris(4-aminophenyl)benzene and terephthaldehyde,[31] and TpAzo-COF made from 1,3,5-triformylphloroglucinol and 4,4'-azodianiline,[32] as light-driven microswimmer examples, in order to explore the microrobotic possibilities for this class of materials and to establish versatile applications. We describe their light-controlled propulsion in biological media, their biocompatibility, as well as their uptake and release of drugs that can be physisorbed to the pores of the material.[25, 33] Since these two COFs have distinct structures and morphologies, we derive design guidelines for their propulsion and cargo-related functions. For theranostic delivery functions of microrobots, we used doxorubicin, insulin, and indocyanine green (ICG), which covers the breath of small molecules and peptides used as therapeutic and imaging agents. We further image the motion of the COFs in real-time and potential in-vivo conditions using optical coherence tomography as well as photoacoustic imaging of COF particle swarms loaded with a near-infrared active dye (ICG). The water-soluble or insoluble drugs and the contrast agents can be loaded to track their release, allowing for first insights into the action of porous drug carriers in real-time clinical imaging modalities. In this way, we designed and investigated in detail the first photoactive intraocular drug carriers for various theranostic applications.

**Results and Discussion**

## COF synthesis and characterization

The COF structures to be used as microswimmers were selected based on their structural and optical properties and synthesized akin to a procedure reported earlier.[31, 32] In brief, the TAPB-PDA-COF nanospheres were obtained by using TAPB (1,3,5-tris(4-aminophenyl)benzene) and PDA (terephthaldehyde) as building blocks, with acetonitrile and Sc(Otf)$_3$ as solvent and catalyst, respectively (see Materials & Methods section for more details) to yield a two-dimensional (2D), imide linked organic network (**Fig. 1a**). These 2D sheets are stacked on each together forming an ordered 3D structure with hexagonal pore channels of a diameter of approx. 3.4 nm, as reported earlier and confirmed for the here modified synthesis by nitrogen sorption, powder XRD, and FT-IR analysis (**Fig. 1b** and **Fig. S1**).[34] The obtained COF submicron particles (henceforth called nanoparticles for better discrimination) have an almost perfectly spherical shape (**Fig. 1c**), which is beneficial for propulsion in fluids.[35] The synthesis yields a very homogeneous product with a Brunauer, Emmett, and Teller (BET) surface area as high as 685 m²/g (**Fig. S1c**) and a narrow size distribution of approx. 452 ± 74 nm (**Fig. S2a,b**). TEM analysis reveals that these nanoparticles consist of agglomerated individual crystallites with sizes between 50 and 100 nm (**Fig. S1c**).

The TpAzo-COF presented here for comparison is synthesized by solvothermal condensation between 1,3,5-triformylphloroglucinol (Tp) and 4,4´-azodianiline (Azo), forming a tautomeric ketoenamine COF (**Fig. 1d**). A highly crystalline product with a 2D molecular structure is obtained (see **Fig. S3** for structural analysis), with slightly smaller structural pores of 2.6 nm and a similar BET surface area of 635 m²/g (**Fig. 1e**). In contrast to the first COF, the particle morphology is much less defined, leaving open large textural voids reminiscent of a sponge (**Fig. 1f** and **Fig. S4**). The overall particle sizes are broadly distributed (7 ± 18 µm), hence being much larger. The primary crystallites forming the COF particle are only 20 nm (**Fig. S4d**), and the TpAzo-COF particles appear to be agglomerates of those.

### Light-induced swimming in aqueous media

To enable light-triggered propulsion, we first investigate the light absorption properties. UV-Vis spectroscopy and Kubelka-Munck analysis show that TAPB-PDA-COF and TpAzo-COF have an optical band gap of 464 nm (**Fig. 2a**) and 616 nm (**Fig. 2f**), respectively with a small absorption tail commonly arising from defect states. Hence, visible light propulsion can be extended up to a wavelength of ~470 nm (blue light) for the small TABP-PDA particles, and to green or even red parts of the spectrum for the large TpAzo-COF. Their light-induced propulsion was studied under a microscope in a microfluidic chamber under ambient conditions to test the phototaxis capabilities while diluting them to 100 µg/ml. First, we focus on propulsion in distilled water. While in the dark, COF particles show only local Brownian motion with a mean displacement speed of 4.5 µm/s for the TAPB-PDA and 3.7 µm/s for the TpAzo-COF, respectively (**Fig. 2b,g**, dashed line). When light from the photodiode is focused on the microswimmers through the microscope, their propulsion speed is significantly enhanced and becomes ballistic, as seen in **Video S1.** The particles move towards the center of the light, then upwards. This way, the light-driven collective assembly or trapping of the microswimmers is made possible. UV excitation at 385 nm propels the TABP-PDA-COF with 13.2 ± 2.4 µm/s, while 470 nm blue light gives an even increased speed of 16.4 ± 3.1 µm/s (~36 bodylengths/s (BLPS)). At 510 nm illumination, no light-enhanced swimming was observed, consistent with the absorption spectrum (**Fig. 2b**). The Tp-Azo-COF is propelled with 4.9 ± 1.2, 12.1 ± 2.1 (~2 BLPS), 8.2 ± 1.7, and 4.2 ± 1.2 µm/s at 385, 470, 560 nm, and 630 nm, respectively (**Fig. 2g**). UV and red light hence do not increase propulsion significantly below Brownian motion and the absolute propulsion speed is lower, but it can be triggered even by yellow light (560 nm).

## Ion tolerance for light-driven microswimmers and phototaxis behavior

Ionic conditions represent a major challenge for light-driven microswimmers.[36, 37, 38] The presence of pores, both structural and textural (=morphological), was suggested previously to enable the propulsion of microswimmers in ionic environments, which includes most of the biological fluids and cell culture media.[11] To confirm this and to widen the insights from different and better controlled structural features present in our model COFs, we first tested them in increasing concentrations of salt (NaCl), see **Fig. 2c,h**.

When propelled at 470 nm, the TABP-PDA-COF does not decrease the speed compared to distilled water up to concentrations as large as 1000 mM. Therefore, the ionic concentration in the media at which the microswimmers' speed is halved (EI50) cannot be attributed.[39] Also, we observe a slightly increasing propulsion speed between 0.5 and 10 mM, with a maximum value of 25.2 ± 3.7 µm/s (54% increase vs. distilled water, 0 mM) at 1 mM NaCl (**Fig. 2c**). An explanation for this non-linear behavior remains to be found. Ionic interactions can be influencing the Helmholtz and Debye layers, as well as the materials' inner space charge layer. As such, light-induced charge carrier stability or recombination will also be affected. On the other hand, the chlorine evolution reaction due to dissolved NaCl may another photocatalytic pathway possibly increasing the reaction rate, and thereby the propulsion speed.[11, 40, 41, 42] These factors currently cannot be studied or disentangled on such size and complex reaction interface. However, their propulsion speed even surpasses our previously reported PHI microswimmers, the only reported system with comparable ionic tolerance.[11]

Similarly, for the TpAzo-COFs, an increased propulsion speed compared to pure water is observed in all ionic conditions (1-1000 mM) at 560 nm illumination, peaking at 1 mM (14.7 ± 2.7 µm/s, 79% increase vs. distilled water) and followed by a 28% relative decay to 10.5 µm/s at 1000 mM. When increasing the wavelength, no active propulsion is observed for TABP-PDA-COF (**Fig. 2b**), but the Tp-AZO-COF exhibits slightly enhanced propulsion even at 630 nm (4.2 µm/s) (**Fig. 2g**).

Next, three standard biological media are studied, namely, Dulbecco's phosphate-buffered saline (dPBS), minimum essential medium (MEM), and MEM plus fetal bovine serum (FBS) (**Fig. 2d,i**), which slightly differ in their components: dPBS contains NaCl, KCl, $Na_2HPO_4$, and $KH_2PO_4$ at ca. 10 g/L (~150 mM) in total; MEM contains the same components as dPBS and additional two amino acids, vitamins, and glucose, some of which can be redox-active agents that help extract not only electrons but especially holes from the microswimmers under illumination to power them.[11, 43] FBS, slightly more viscous, adds nutrients for cell growth and imitates the conditions found within the body.[11] At 470 nm illumination, the mean speeds of the TAPB-PDA-COF microswimmers in dPBS, MEM, and MEM + FBS are 20.4 ± 4.5 µm/s, 20.9 ± 2.8µm/s and 17.6 ± 3.3 µm/s, respectively. These speeds are again higher than in distilled water (16.4 ± 3.1 µs/s). The slight decrease upon FBS addition can be attributed to the increasing viscosity or other surface interactions with the proteins present in the FBS.

Very similar behavior is observed with the TpAzo-COF at 560 nm, where the swimming speeds are equivalent to the maximum value in 1mM NaCl, or even slightly higher (13.8 ± 3.4 µm/s, 16.2± 3.4 µm/s, 14.7 ± 3.6 µm/s, and 11.8 ± 2.2 µm/s in dPBS, MEM (with and without glucose), and MEM + FBS). A difference however is observed when glucose, a well-oxidizable fuel,[11, 43] is absent – the speed is reduced. Its vital role as fuel for propulsion is clearly visible when illuminating TpAzo at 630 nm in MEM that contains glucose, where efficient propulsion, independent of FBS, is observed (10 ± 2.7 µm/s and 7.4 ± 1.8 µm/s respectively). This purely red light-induced photocatalytic motion in the presence of high ion concentrations and without using potent and toxic fuels is unprecedented.[29, 44] However, the still efficient propulsion at 560 nm without glucose in MEM confirms that the other ingredients (including dissolved oxygen[11, 19]) may also assist motion induced by photocatalysis, or at least do not hamper it. These experiments not only show the superiority in performance over current inorganic microswimmers in high-salinity media but also highlight how crucial facile redox species are that can act as fuel for propulsion, akin to photocatalysis in general, and especially if sub-band gap trap states might be partially involved (630 nm illumination).[43, 45, 46] Such a substantial shift toward the red part of the spectrum that can penetrate deeper tissues makes organic and small band gap microswimmers (especially with trap states in the gap) attractive for micromachines not just in-vitro, but even for in-vivo conditions.

Light-driven directional propulsion control

Phototaxis is the property by which microswimmers swim towards or away from the direction of incident light (i.e., positive or negative phototaxis), which often depends on their surface charge.[47, 48] It enables direction control, opposite to random ballistic displacement usually observed with Janus particles.[11, 49] When the COF microswimmers were illuminated by a directed light source from the side with a 45° angle, both TABP-PDA-COF and TpAzo-COF microswimmers exhibit positive phototaxis, and swim toward the light that can propel them (**Fig. 2 e, j, and video S2**). TABP-PDA-COF and TpAzo-COF particles move with mean speeds of 13.3 ± 1.8 µm/s and 7.6 ± 0.8 µm/s, at 470 nm and 630 nm illumination in water and MEM, respectively. This apparent increase in the particle speed compared to vertical illumination could be attributed to the larger parallel component of the light direction to the propulsion direction when the samples were illuminated from the side. When the samples are illuminated from the bottom, only the side-wise motion component is measured as a common standard, artificially decreasing the actual velocity.[50, 51] Similar findings have been found on carbon nitride microswimmers, which were discussed in more detail in our previous study.[11] The required symmetry breaking is created by the side-wise illumination and, thereby, an artificially created Janus structure results from the self-shadowing of the microswimmers.[13, 47]

Biocompatibility of COFs

In order to be used in potential biomedical applications and to ascertain biocompatibility, microswimmers should have no significant cytotoxicity. Hence, we tested the cytotoxicity of the microswimmers with human umbilical vein endothelial cells (HUVEC) in dMEM with FBS. Different concentrations of TAPB-PDA-COF and TpAzo-COF microswimmers (3.1-25 µg/ml) were incubated with HUVECs in the dark, and their viability was investigated with calcein-based live/dead fluorescence staining of the cells after 24 hours. The cells with TABP-PDA COF were completely viable, and they did not show any significant decrease in viability even at high concentrations, both with illumination and without illumination at 470 nm with maximum light intensity, 10 mW/cm$^2$, for 30 minutes), as seen in **Fig. 3 a,** which is visible also in live cell fluorescent images in **Fig. 3 b**. TpAzo-COF (**Fig. 3 c,d**) shows lower cell viability in comparison with TAPB-PDA-COF, with 93% and 75% HUVEC cell viability in 25 µg/ml concentration (in dark and with 630 nm illumination (10 mW/cm$^2$), respectively). Also, at concentrations of 3.1 µg/ml, the viability is decreased to 88% in comparison to the TABP-PDA COF. However, this fairly good viability indicates that also the TpAZo COF can be used at lower concentrations for drug delivery applications. Generally, illumination seems not to affect the viability at low concentrations (3.1 and 6.25 µg/ml), and only slightly at 12.5 and 25 µg/ml for both COFs. These results also suggest that light-induced propulsion induces only minimal cytotoxicity in the range of light-driven propulsion periods. Compared to carbon nitride microswimmers, which have a larger band gap (2.5 eV, 450 nm) and a very low-lying valance band, and therefore enable more redox reactions with organic matter, including cells in principle, the use of 470 nm or 630 nm light with our TABP-PDA COFs and TpAzo COFs shows potential for reduced cell death [with 97% and 88% cell viability after 30 minutes of light in 3.1 µg/ml concentrations of TAPB-PDA-COFs and TpAzo-COFs, respectively] and makes especially the TABP-PDA COFs more applicable to practical applications such as drug delivery.[11] A previous study with primary cells from mouse splenocytes further confirmed no detectable level of IL-12 (a pro-inflammatory cytokine) in the untreated samples in concentrations used above in the dark.[52]

### Drug loading, drug delivery, and hyperthermia

To explore the COF microswimmer's applicability to biological environments, we also studied their potential as drug carriers with different pharmacological agents. The differently pronounced textural and structural porosity of the TABP-PDA and TpAzo-COFs (see **Fig. 1** and **Fig. S1-S4**), which enables ionic tolerance (**Fig. 2c,h**), is not only beneficial for motion but also as space to take up, transport and deliver therapeutic drugs. We studied and compared how the structural features enable interactions with such cargo in the following experiments. For this reason, we chose an imaging agent, indocyanine green (ICG), and two different pharmacological agents with different Biopharmaceutics Classification System (BCS) classes: doxorubicin (DOX) (Class III) and insulin (Class I).[53] Also both pharmacological agents are currently used to treat common ocular disorders.[54]

First, we tested the loading of DOX, a chemotherapeutic agent against various cancer types, including retinoblastoma.[55] 200 µg of DOX was added to a suspension of 100 µg of COF microswimmers dispersed in 1 mL MEM, resulting in 138 µg DOX encapsulated (loading efficiency of 138%) on the TABP-PDA-COF microswimmers after 24 hours, and 75% for TpAzo-COF. Due to the small molecular size of DOX (~1.1 nm approximate molecular diameter), the molecule should fit into the structural pores of both COF structures (3.4 nm and 2.5 nm), while adsorbing also on the inner textural surface. The overall negative surface charge on both COF microswimmers attracts the positively charged DOX molecules in physiological pH values and gives rise to stable loading. Since the overall surface areas are similar within 10%, it appears that differences in polarity or hydrogen bonding, possibly mediated by the carbonyl groups of TpAzo-COF, enable electrostatic repulsions with the DOX molecules and interfere with DOX uptake in TpAzo-COF structures, which is also correlated with the zeta potential measurements. While the positive zeta potential of the TABP-PDA-COF ($\zeta_{TABP-PDA-COF}$ = 12.13 ± 1.28 mV) reduces agglomeration and enables sufficient drug loading values, the negative zeta potential of the TpAzo-COF ($\zeta_{TpAzo-COF}$ = -19.67 ± 0.68 mV) leads to agglomerations and reduces drug loading due to electrostatic repulsions.[56] In addition, a lower crystallinity and thereby, possibly decreased accessible pore volume of TpAzo-COF are expected to lead to reduced DOX uptake. Overall, the DOX uptake of both COF materials is among the highest reported, relative to other artificial structures using physical encapsulation.[11, 57]

The DOX release can be achieved by changing the pH to slightly more acidic conditions, i.e., from pH = 7.2 to 5.5 (**Fig. 3 e, g**), which is achieved by adding HCl to PBS. The TABP-PDA microswimmers release 95 µg of DOX within 60 minutes, which is significantly boosted compared to the weak, passive release also observed (12 µg). The passive release is commonly observed when drugs such as DOX are not entirely trapped or encapsulated within porous structures but physisorbed to the surface. Encapsulation within the TpAzo-COF, with a more open texture, appears more stable, as evidenced by the lower passive release at pH 7.2 (5 µg in 60 minutes). In line, a reduction of pH to 5 only releases 7% in 60 min, whereas a pH 3.5 yields 25% and is more reasonable as a release trigger. The acid-triggered DOX release in the TABP-PDA-COF and TpAzo-COF microswimmers can be seen in fluorescence imaging in **Figure 3f,h**, respectively. The enhanced drug delivery of microswimmers at lower pH has the potential to enable the targeted therapy in tumor or infection environments, which typically have acidic pHs.[58, 59]

We also studied the loading and release of peptide (insulin), a frequently used drug in diabetic retinopathy and convenient for light-controlled drug release applications.[60, 61] Its larger molecular size of ~3 nm makes larger pore sizes on the COFs desirable to allow for an efficiently encapsulated loading. Indeed, insulin loading was observed on both COFS, 60% for TABP-PDA-COF (3.5 nm pore size] and 40% for TpAzo-COF (2.5 nm pore size) (**Fig. 3i,k**), which suggests that physisorption of the drugs occurs on the outer surface of the textural pores and that the structural pores can assist stable uptake.

Similar to DOX release from the COF structures, changing pH enables insulin release from both COFs. While the TABP-PDA-COF shows a continuously increasing cumulative release of approximately 35 µg/ml within 60 min at pH 5 already, which may be desirable for slower dosing, the TpAzo-COF releases its cargo rather instantly (within 10 min), and at lower amounts (~10 µg/ml in more acidic pH 3.3 again). With both drugs, no visible light-triggered release was observed, opposite to the carbon nitride systems reported earlier with DOX. However, as seen herein, the absence of such a property can be very beneficial since it enables the decoupling of motion control and drug release, which would otherwise have to co-occur.[11]

As a third theranostic agent to load onto COFs, we used ICG dye, commonly used in diagnosing retinal diseases.[62] Firstly, we investigated ICG loading and near-infrared laser-induced hyperthermia capabilities; then, we focused on medical imaging of ICG-loaded COF microswimmers with photoacoustic imaging and optical coherence tomography. As is the case with drug delivery, TABP-PDA-COF has a pore size larger than the size of the ICG (~2.9 nm molecular diameter on its longest axis); hence, the drug is presumably loaded better into the structural pores of the TABP-PDA COF (3.5 nm), while in the case of TpAzo-COF, it appears to dominantly bond to the bigger, textural pores **(Fig. 4a)**. After ICG was loaded onto the both, TpAzo-COF and TAPB-PDA-COF microswimmers at two different loading levels (50% and 100%, w/w), they were irradiated with a near-infrared (NIR) laser at 808 nm.[63] ICG-loaded TABP-PDA-COFs achieved quick heating to 66 °C and 69 °C after only 3 minutes of 808 nm NIR irradiation for 50% and 100% loading, respectively. Compared to TABP-PDA-COFs, ICG-loaded TpAzo-COFs heated up to 42 °C and 45 °C for 50% and 100% loading under the same NIR illumination conditions **(Fig 4 b, c)**. Heat generation and accumulation are always affected by heat transport to the environment. Assuming similar absorption and hence heat generation at the same loadings, these findings indicate that indeed, ICG transfers the heat slightly better by binding to TABP-PDA COF, and that the TpAzo COF dissipates accumulated heat faster to the environment due to its more open shape, and thereby reaches lower temperatures over extended times. In both cases, this NIR-controlled hyperthermia behavior of both COFs could be helpful for novel intraocular photodynamic therapy application, which is already in the clinical trial phase for ICG dye.[64] Compared to other novel intraocular photothermal therapy agents in the recent literature, especially TABP-PDA-COFs with pores enabling ICG uptake into the material's structural pores and intense heating from 25 °C to 69 °C in 3 minutes, shows significant potential for the photodynamic combined therapy applications that are used to degrade cells by heat generation.[65, 66]

## Photoacoustic imaging and optical coherence tomography

Imaging microswimmers as they move in different fluids is one of the most critical enablers for their potential in vivo applications.[67] For this purpose, we selected to study two clinical imaging methods: optical coherence tomography (OCT) and photoacoustic (PA) imaging. OCT is the gold standard high-resolution clinical imaging method to observe intraocular structures and is accessible in most ophthalmology clinics worldwide.[68] PA is an emerging imaging technique that combines the resolution of optical imaging with the depth of penetration of ultrasound imaging. In recent years, PA has been used in ophthalmology as it shows significant advantages in imaging deep ocular structures, such as lymphatic drainage and choroidal vasculature.[69, 70] While in the PA imaging method ICG was used as a contrast agent to enhance the visualization of COF microswimmers in the complex environment of intraocular fluids, COFs were imaged in intraocular structures and ocular fluids without any contrast agent during OCT imaging. Different concentrations of ICG are loaded onto the COF microswimmers and imaged under photoacoustic imaging **(Fig. 4d,e)**. While TAPB-PDA-COFs achieve up to 500 mean pixel intensity (MPI) at 815 nm, which is the highest peak in the emission spectrum of ICG, TpAzo-COFs achieve 250 MPI under the same imaging conditions. These signal intensity increases correlate with the concentration of the ICG in the COF loading suspension and also the drug uptake ability of both COFs, which correlate with other drug loading experiments. Imaging uptake and delivery of therapeutic agents on microswimmers will be helpful in the targeted in vivo drug delivery experiments.[71]

As a next step, the light-driven propulsion of the COF microswimmers in intraocular fluids was observed using PA imaging. In both vitreous and aqueous fluids, COFs were illuminated in the same fashion as in the light-induced swimming experiments in various media and then observed with photoacoustic imaging for 30 minutes (**Fig. 5a-c**). Except for TpAzo-COF in vitreous humor under 630 nm light illumination, an increased ICG emission signal was observed in the focus areas for all experimental groups. These results indicate that the light-driven collective motion of both COF microswimmer types could be trackable under PA imaging.

For clinical applicability, we observed and measured the light-driven swimming of COF microswimmers in intraocular fluids under real-time OCT. While the mean speeds of the smaller and spherical TAPB-PDA-COFs were 12.1 ± 1.7 µm/s in aqueous humor and 7.6 ± 0.8 µm/s (~16.8 BLPS) in the vitreous humor, mean speeds of TpAzo-COFs were slightly increased to 14.2 ± 1.5 µm/s in aqueous humor and 8.8 ± 1.0 µm/s (~1.25 BLPS) in the vitreous humor under 470 nm light illumination (**Fig. 5d and Video S3**). Compared to the previous intraocular microrobotic studies employing magnetic actuation of helical microswimmers, the speed of the microswimmers in terms of BLPS was significantly higher, ~16.8 BLPS in the current study vs. ~5.3 BLPS for the fastest magnetic intraocular microswimmers previously.[23] Light-driven accumulation behavior of both COF microswimmer types in the focus of the light was trackable under real-time OCT imaging without any contrast agent loading (**Fig. 5e and Video S4**). Additionally, their light-driven propulsion in 470 nm wavelength light was also trackable even inside an ex vivo porcine eye with anterior segment OCT imaging (**Video S5**). The COF-based microswimmers are the first intraocular microswimmers that can swim and be trackable inside the eye without any contrast agent or surface modification. TpAzo-COFs were actuated faster, opposite to the previous experiments, which highlights that a perfectly spherical shape of TAPB-PDA-COFs alone is not of dominating benefit for mesh-like heterogeneous structures. Although the reasons for this inverted swimming speed remain to be clarified and likely depend on photocatalytic reaction rates in the respective environment, it is possibly also linked to the increased viscosity and fibrillary mesh structures in the aqueous and vitreous humor that overall decrease the propulsion speed of both COF microswimmer types compared with previous aqueous conditions.[30] These results show that both COF microswimmer types are suitable microrobotic drug delivery agents under both PA and OCT imaging, while enabling actual biomedical applications inside body fluids, especially for intraocular structures. With the help of their promising drug delivery and NIR-based hyperthermia abilities, they could solve the active retinal drug delivery problems in various ocular disorders.[54] They could be easily loaded with DOX for chemotherapy without adverse effects on retinoblastoma patients or with insulin to treat increased ocular pressure.[72, 73] COF-based microswimmers can easily be controllable with visible light, instead of other passive nanomedicine agents in ophthalmology clinics and they do not require complex and unalterable magnetic coil setups with narrow working spaces.[21, 23]

**Conclusion and Outlook**

In this manuscript, we have studied two structurally and texturally distinct COF microswimmer types with tunable nanopore sizes towards their potential intraocular medical applications as multifunctional microswimmers. This comparison of COFs from two different families with distinct morphologies and drug loading capabilities yielded promising results in terms of biocompatibility, imaging, drug delivery, and visible light-induced propulsion in ionic and biological media, surpassing the applicability of current magnetically actuated microswimmer-based systems – without a need of further structural modification or sophisticated structural engineering. Simultaneously, the COF microswimmers can be propelled by visible and even red light in ionic and biological conditions (**Fig. 2**). Although some medium-dependent propulsion trends at low salt concentrations remain to be clarified, their porous structure, coupled with photocatalytic activity, seems key to efficient photocatalytic motion without dedicated toxic fuels or harm to the tissue. A compact spherical shape, as achieved by the size-modified synthesis of the TABP-PDA COFs, appears beneficial for fast propulsion, enabling bubble-free motion at 36 BLPS while opening up possibilities for mobility in the intraocular region. On the other hand, large and texturally more porous structures, as observed for the TpAZo-COF, enable similar absolute propulsion speeds in ionic conditions, albeit at a much-reduced speed relative to their size (~2 BLPS). The explanation for this behavior remains to be found and rationalized by numerical models, especially since simple fluid dynamics and the applicability of Reynolds numbers, which do not include inner flow, are not suited for these systems.[74] Both microswimmers allow for precise motion control as single particles by their phototactic properties, enabling complex curvilinear navigation around obstacles in principle and collective motion for particle (re-)assembly (**Fig. 2**).[11, 19, 75] We show that large structural and textural pores enable the loading of different drugs and dyes (e.g., insulin, ICG, and DOX) but that the pore size itself only plays a partial role in (stable) uptake, since textural surface area also contributes to drug binding, as clearly visible by the different uptake properties of the small drug DOX (**Fig. 3**), whereas larger molecules, such as insulin or ICG, can stay more stably bound, even in lower loading amounts (**Fig. 4**). Since the drug binding and release is also affected by chemical interactions between the COF backbone and the drug, independent of pore size and surface area, future material design should focus on optimizing these interaction factors to broaden our insights.

The versatility of COFs, not only on the morphological but especially on a molecular level, is anticipated to enable tailored approaches to tune the adsorption and desorption properties of drugs, akin to their use on gas sorption.[76] Modifications of these interactions, especially by external stimuli, such as pH changes, light, viscosity changes, and oxygen content in the vicinity, can enable the desired interaction strength with the cargo and its release kinetics.[11] This possibility is anticipated to enable tailored, targeted, and especially semi-autonomous therapy not only for in vitro but also for in vivo applications.[77, 78]

We further demonstrated medical imaging of the ICG-loaded COFs, enabled by photoacoustic imaging and optical coherence tomography. In principle, both of them enable the visualization of swarms and motion of large individual particles, providing more detailed insights into local propulsion and release properties inside the eye or soft tissues where visible light cannot penetrate easily. Since the ICG loading can be kept very low in the porous COFs while maintaining a high signal intensity **(Fig. 4)**. Optical coherence tomography inside eye tissue also enables real-time imaging studies of drug-loaded microswimmers and evaluation in intraocular fluids and structures, laying the grounds for a more detailed understanding of release properties and burst kinetics for various theranostic agents. By decoupling COF microswimmers' motion control and release mechanism, a broad range of independent functionalities is made possible on these porous organic structures in parallel. We anticipate that especially simultaneous imaging, drug release, and NIR light-assisted photothermal therapy capabilities will offer additional theranostic abilities beyond what current state-of-art noninvasive photodynamic therapy techniques could achieve.[79] In the near future, they could be functionalized in ophthalmology clinics for multimodal therapy and imaging of retinal diseases, such as retinoblastoma, diabetic retinopathy, or glaucoma.

**Materials & Methods**

*Synthesis and preparation of covalent organic frameworks*

Synthesis of TAPB-PDA-COF was carried out according to a previous report with minor changes.[34] In a typical colloidal reaction, 1,3,5- tris(4-aminophenyl)benzene (TAPB) (0.030 mmol, 10.4 mg) and terephthaldehyde (PDA) (0.044 mmol, 5.96 mg) were dissolved in 14 mL acetonitrile. After 10 minutes of sonication, a solution of Sc(OTf)$_3$ (0.014 mmol, 7.00 mg) in 7 mL acetonitrile was added dropwise at room temperature under slight stirring. After 24 hours of reaction, the solvent was exchanged for distilled water by centrifugation for five times (795 g for 10 minutes each). For solids characterization, the particles were precipitated by adding 0.5 mL of 1 M NaCl solution, washed with methanol, and dried by supercritical $CO_2$ on a Leica EM CPD300 instrument. TpAzo-COF was synthesized according to a previous report.[80]

*Brunauer–Emmett–Teller (BET) measurements and analysis*

Nitrogen sorption measurements were performed on a Quantachrome Instruments Autosorb iQ MP at 77 K. Before the gas adsorption studies, the samples were degassed for 12 h at 120 °C under a vacuum. Multipoint BET surface area calculations and pressure ranges were chosen according to the linear region on the BET plot in the range between 0.05 and 0.35 $P/P_0$. Pore size distribution was determined from Nitrogen adsorption isotherms using the NLDFT cylindrical pores in the carbon model for nitrogen at 77 K.

*PXRD measurements and analysis*

Powder X-ray diffraction experiments were performed on a Stoe Stadi P diffractometer (Cu-K$_{\alpha 1}$, Ge(111) in Debye-Scherrer geometry. The samples were measured in sealed glass capillaries (OD = 1.0 mm) and spun for improved particle statistics.

*Transmission electron microscopy (TEM) and scanning electron microscopy (SEM)*

Transmission electron microscopy was performed with a Philips CM30 ST (300kV, LaB$_6$ cathode). The samples were prepared dry onto a copper lacey carbon grid (Plano). Images were recorded with a TVIPS TemCam-F216 CMOS camera. The program EM-Menu 4.0 Extended was used for analysis.

SEM images were obtained on a Zeiss Merlin or a VEGA TS 5130MM (TESCAN) with an InLens detector using electron energy of 1.5 kV. The samples were cast on indium-doped tin oxide (ITO) substrates, and a 3 nm-thick iridium film was sputtered on them to reduce charging.

*UV-VIS measurements and analysis*

For diffuse reflectance UV–visible absorption, spectra were collected on a Cary 5000 spectrometer (referenced to barium sulfate). Absorption spectra were calculated from the reflectance data using the Kubelka-Munk and assuming a direct band gap.[81]

*Zeta potential measurements*

The Z potential was determined using a Malvern nano Zs zetasizer. Dispersions of 0.5 mg/mL COF in 10 mM aqueous NaCl were sonicated 15 min before zeta potential experiments. Surface charge values represent the mean of 3 experiments and their standard deviation is indicated.

*Light-driven propulsion experiments*

The spectral irradiance of the illumination in the microscope was measured at the place of the sample chamber with a calibrated Ocean Optics OCEAN-FX-XR1-ES spectrophotometer after attenuation by a neutral density filter. The results have been normalized to the filter attenuation and the spot size of the light beam in the microscope. It was measured to be 2.0 ± 0.5 mm in diameter, resulting in a relative experimental error of 50% after the error propagation calculation. In the case of visible light propulsion, a broad-spectrum low-intensity white LED is illuminated from the top, and lights with various wavelengths (385 nm, 470 nm, 510 nm, 560 nm, and 630 nm) are illuminated through the microscope objective. The intensity of the microscope light (1 mW/cm$^2$ for the control experiments in the dark and 2 mW/cm$^2$ for imaging during UV light-based propulsion) was increased to 10 mW/cm$^2$ for visible light propulsion. For photocatalytic and PEC experiments, a calibrated Thorlabs S425C/PM100D optical power meter directly measured the light intensity.[19] All light intensities are used in the light propulsion experiments under the ocular safety limit (54 mW/cm$^2$) for ophthalmic devices.[82]

*Biocompatibility experiments*

Human umbilical vein endothelial cells (CRL-1730 [HUVEC], ATCC, Manassas, VA) were grown in dMEM supplemented with 10% (v/v) FBS and 1% (v/v) penicillin/streptomycin (Gibco, Grand Island, NY, USA) at 37°C in a 5% CO$_2$, 95% air-humidified atmosphere. Cells were reseeded after growing to confluence into μ-Slide eight-well plates (Ibidi GmbH, Gräfelfing, Germany) at a cell density of 25 x 10$^3$ cells/well and incubated for two days. HUVEC cells were incubated with TAPB-PDA or TpAzo COF microswimmers at varying concentrations (3.1 to 25 μg/ml) for cytotoxicity testing. Then, the cell viability was measured using a LIVE/DEAD assay (Thermo Fisher Scientific, Waltham, MA) incorporating calcein-AM (green) and ethidium homodimer-1 (red) dyes. After 24 hours of incubation with the COF microswimmers, live-dead cell numbers were calculated from fluorescence microscopy images. Furthermore, cytotoxicity of microswimmers during light actuation (470 nm for TAPB-PDA and 630 nm for TpAzo, 10 mW/cm$^2$ and 4 mW/cm$^2$, respectively) was tested by live/dead staining of HUVEC cells right after and 24 hours after actuation of COF microswimmers for 30 minutes.[11]

### Drug & ICG loading and release tests

The loading efficiency was measured by centrifuging the DOX (44583, Sigma-Aldrich, St. Louis, USA) or insulin (I3661, Sigma-Aldrich, St. Louis, USA) loaded microswimmers and comparing the optical density (OD) of the supernatant with the precalibrated OD of DOX or insulin (200 µg/ml) at 480 nm. Both COF microswimmers (100 µg/ml) were dispersed with DOX or insulin (200 µg/ml), and this solution was stirred in the dark for 24 hours to allow the drugs to be adsorbed. After 24 hours, the suspension was centrifuged, and the supernatant was used for measuring the drug loading. The drug-loaded COF solution was washed three times with water and stored in dPBS at +4°C for further delivery experiments. For the pH release, the pH of the resulting HCl-diluted PBS solution was checked using a pH meter to confirm the stability of the pH during the release experiments.[11]

### NIR-based remote heating of ICG-loaded COF particles

TpAzo-COF and TAPB-PDA-COF loaded with 50% and 100% ICG were loaded in microtubes and irradiated with a NIR laser (808 nm, 0.6 W/cm$^2$). Thermal images were obtained, and temperature information was recorded with a thermal infrared camera (ETS320, FLIR Systems).

### Photoacoustic imaging measurements and analysis

The photoacoustic (PA) signal characterizations were performed inside a Multispectral Optoacoustic Tomography device (MSOT 512-element transducer, iThera Medical) system with three scanning steps of 0.2 mm at different wavelengths. The samples with different concentrations were prepared inside a transparent stripe and embedded in an agar phantom (1.5 g/100 mL agar-DI water). The same preparation was done for the control sample. The agar phantom was placed at the center of the transducer arrays. The measurements were then taken for a range of wavelengths (660 – 980 nm), and each image was repeated three times for each laser pulse and then averaged. A circular region of interest (ROI) was chosen for calculating the PA signal at each wavelength. Finally, the diagrams were plotted against the control sample for all concentrations.

For PA imaging of light-induced motion of nanoparticles, a handheld 3D photoacoustic probe (256-element transducer, iThera Medical) was used for real-time tracking. The laser wavelength was set at 800 nm, and the image sequences were taken at 10 frames per second. Then, a volumetric image of 20 × 20 × 20 mm³ was constructed from three orthogonal imaging planes. The real-time change in the signal intensity at the light actuation spot indicated the movement of the nanoparticles.

*Optical coherence tomography (OCT)*

The fresh porcine eyes were purchased from Ulmer Fleisch food factory, Ulm, Germany. Within six hours after the euthanasia of the animals, a set of enucleated eyes stabilized to the holder, and COFs were injected with a 30G syringe in the anterior chambers of the porcine eyes before OCT imaging. Besides that, aqueous humor was removed from another set of fresh porcine eyes with the help of 30G trocar and cannula. For vitreous collection, a classical vitrectomy procedure is followed.[83] The intraocular fluids with COFs were injected into a cylindrical tubing and observed via OCT (TEL320C1 – Spectral Domain OCT System, Thorlabs). The motion inside the leg was recorded with an image speed at a medium sensitivity (76 kHz). The refractive index was set to 1.00, and the Hann filter was used for the apodization window. The A-scan averaging was set to 1, and the B-scan averaging to 1 with a pixel size of 6.5 μm.

**Author contributions**: F.P., V.S., B.V.L., and M.S. conceived and designed the project. F.P., V.S., and E.Y. wrote the manuscript, with input and corrections from all authors. A.R. and L.Y synthesized and characterized the materials. V.S. and X.L. performed the light propulsion experiments and analyzed the data. E.Y. performed and analyzed in vitro biocompatibility tests. X.L. and M.B.A. performed and analyzed drug loading experiments. B. A. performed and analyzed NIR hyperthermia experiments. A.A., E.Y., and P.W. performed and analyzed the photoacoustic imaging. E.Y. isolated porcine intraocular fluids and performed optical coherence tomography. M.S., F.P., and B.V.L. supervised the research. All authors contributed to the discussion of the data and overall results.

**Data availability**: All data are available from the corresponding author upon reasonable request.

**Acknowledgments**: The authors acknowledge Viola Duppel for SEM and TEM image acquisition. We thank Julia Kröger for the fruitful discussions. Support by the Max Planck Society, the Bavarian Research Network SolTech (B.V.L.), and the Deutsche Forschungsgemeinschaft (DFG) via the cluster of excellence "e-conversion" (project number EXC2089/1–390776260) is gratefully acknowledged. F.P. has received and acknowledges UKRI funding under the grant reference EP/X027449/1.  E.Y. has received funding from the European Union's Horizon 2020 research and innovation program under the Marie Skłodowska-Curie grant agreement [PHOTODOCTOR].**References**

1. M.Sitti. Mobile microrobotics. *MIT Press, Cambridge, MA* 2017.

2. Erkoc P, Yasa IC, Ceylan H, Yasa O, Alapan Y, Sitti M. Mobile Microrobots for Active Therapeutic Delivery. *Advanced Therapeutics* 2019, **2**(1): 1800064.

3. Dupont PE, Nelson BJ, Goldfarb M, Hannaford B, Menciassi A, O'Malley MK*, et al.* A decade retrospective of medical robotics research from 2010 to 2020. *Sci Robot* 2021, **6**(60): eabi8017.

4. Wang B, Kostarelos K, Nelson BJ, Zhang L. Trends in Micro-/Nanorobotics: Materials Development, Actuation, Localization, and System Integration for Biomedical Applications. *Adv Mater* 2021, **33**(4): e2002047.

5. Sitti M, Ceylan H, Hu W, Giltinan J, Turan M, Yim S*, et al.* Biomedical Applications of Untethered Mobile Milli/Microrobots. *Proc IEEE Inst Electr Electron Eng* 2015, **103**(2): 205-224.

6. Mujtaba J, Liu J, Dey KK, Li T, Chakraborty R, Xu K*, et al.* Micro-Bio-Chemo-Mechanical-Systems: Micromotors, Microfluidics, and Nanozymes for Biomedical Applications. *Adv Mater* 2021, **33**(22): e2007465.

**Graphical Abstract:**

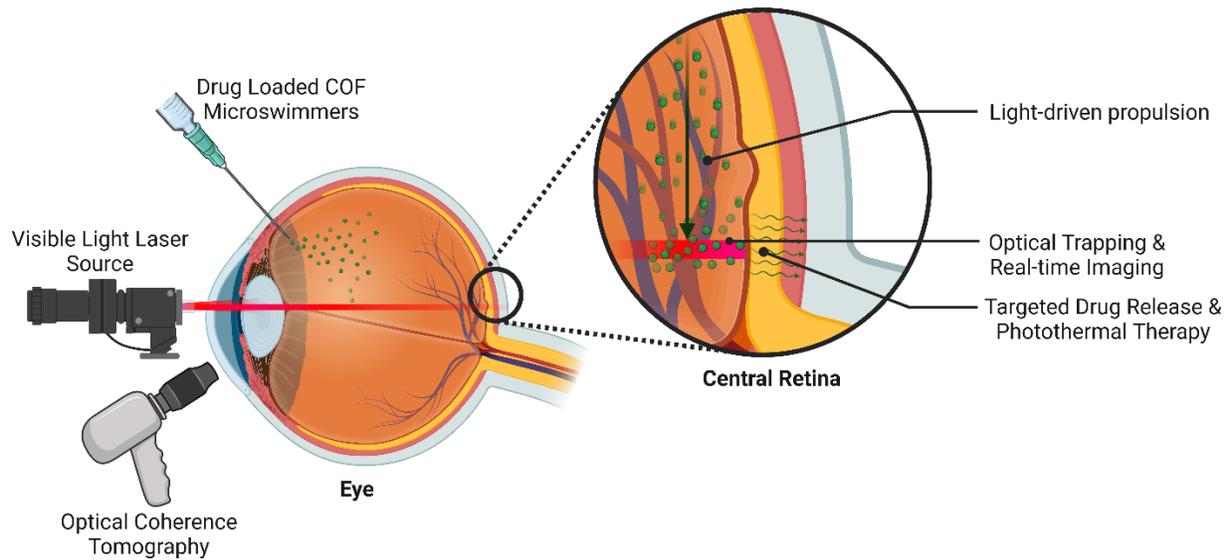

Conceptual illustration of light-driven and light-steered COF microswimmers towards targeted intraocular drug delivery and photothermal therapy applications under optical coherence tomography-based real-time imaging.

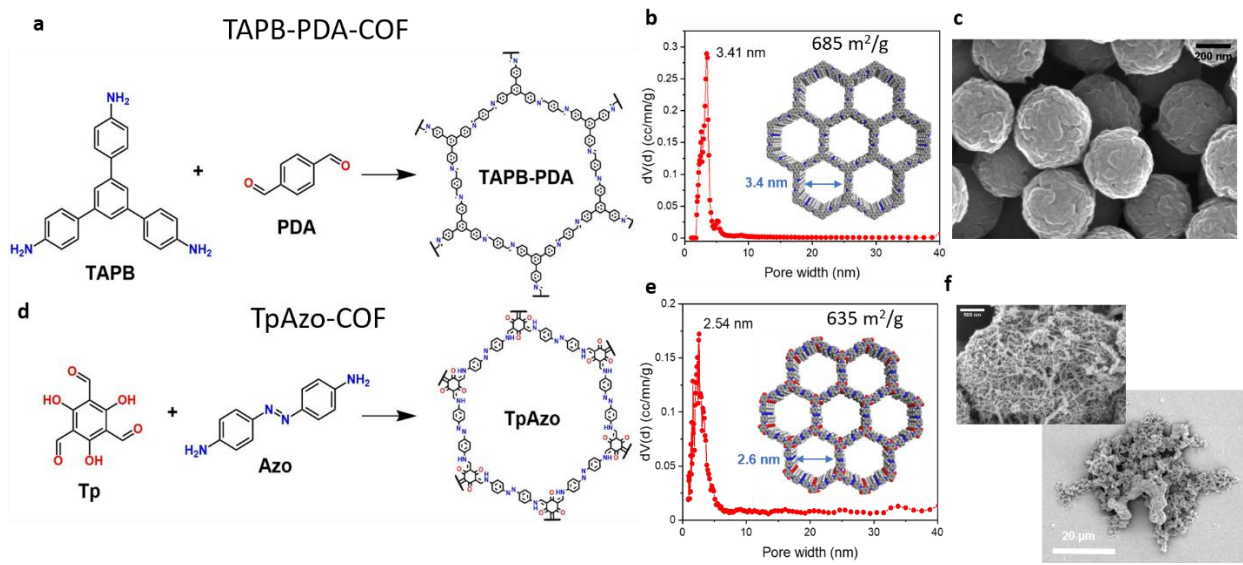

**Figure 1: Structural properties of the two types of COF particles used as light-powered microswimmers. a-c:** Imine-linked TABP-PDA-COF nanoparticles. **a:** Precursors for synthesis and molecular structure of the 2D covalent organic framework that stacks in the third dimension. **b:** Calculated pore size distribution from nitrogen sorption isotherms at 77 K (see **Fig. S1, S2** for details), highlighting a fairly uniform pore diameter of 3.4 nm. **c:** SEM image of TABP-PDA COF nanoparticles with a narrow diameter distribution around 450 nm. **d-f:** Azo-linked TpAzo-COF microparticles. **d:** Precursors for synthesis and molecular structure of the 2D network that stacks in the 3rd dimension. **e:** Calculated pore size distribution from nitrogen sorption isotherms at 77 K (see **Fig. S3, S4** for details), highlighting a relatively uniform pore diameter of 2.6 nm. **f:** SEM images of the TpAzo-COF microparticles with a sponge-like structure and high levels of textural porosity, including macropores and heterogeneous size distribution (6.97 ± 17.62 µm, see **Fig. S3, S4**).

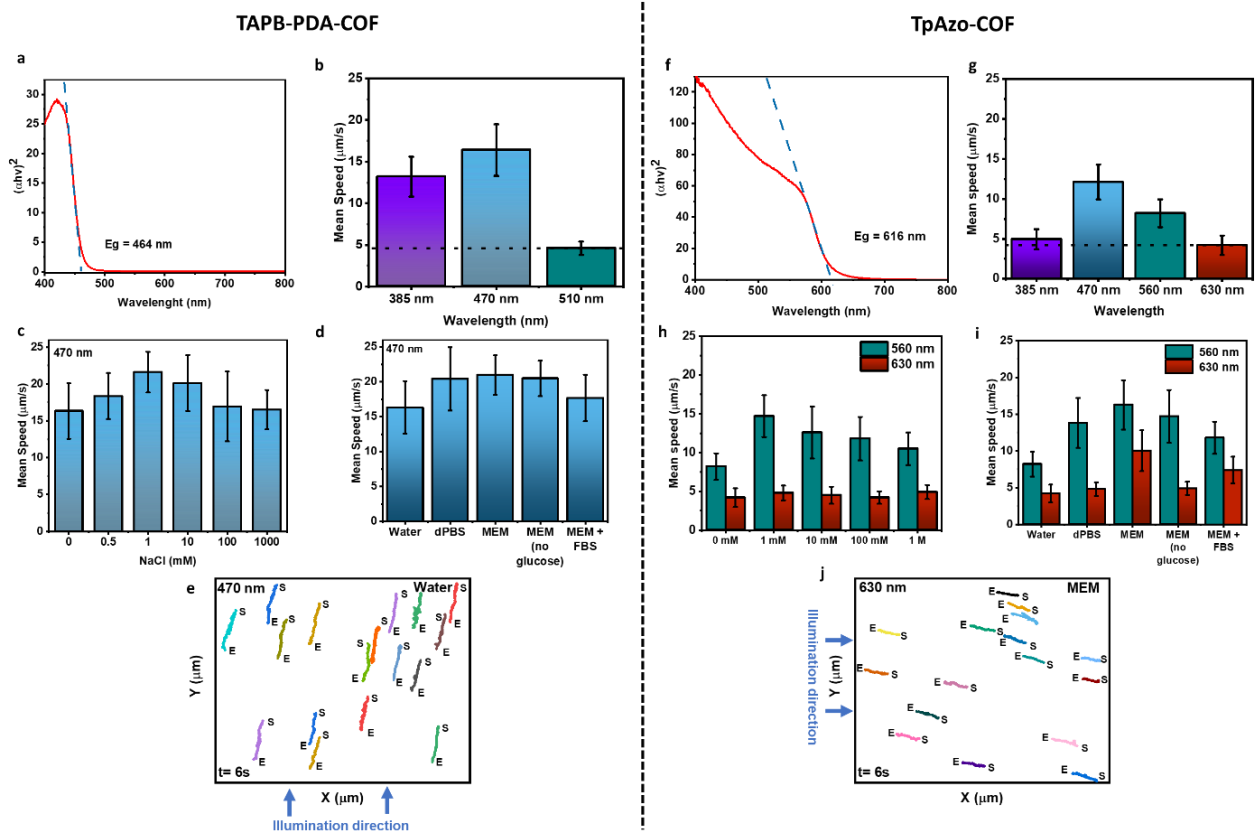

**Figure 2: Optical properties and propulsion of TABP-PDA-COF and TpAzo-COF microswimmers in water and ionic media and their phototaxis behavior. a, f:** Absorbance properties and optical band gap extracted from UV-Vis diffuse reflectance spectra of TABP-PDA-COF (**a**) and TpAzo-COF (**f**) particles, respectively, measured in the solid state. **b, g:** Mean speeds of the COF microswimmers illuminated in distilled water at different wavelengths under the microscope. The dashed line denotes the local Brownian motion speed. Density: 100 µg/ml, N = 50 particles. Error bar = S.D. **c, h:** Propulsion in NaCl with increasing concentration and wavelength highlighting strong ionic tolerance for light-driven propulsion. **d, i:** Comparison of propulsion speed in different commonly used biological media (dPBS, MEM) and MEM modified by removing glucose or adding FBS. Density: 100 µg/ml, N = 50 particles (a-d). Mean ± S.D. **e, j:** Phototactic control of diluted COF microswimmer particles following illumination from the side (S=start, E=end of trajectory).

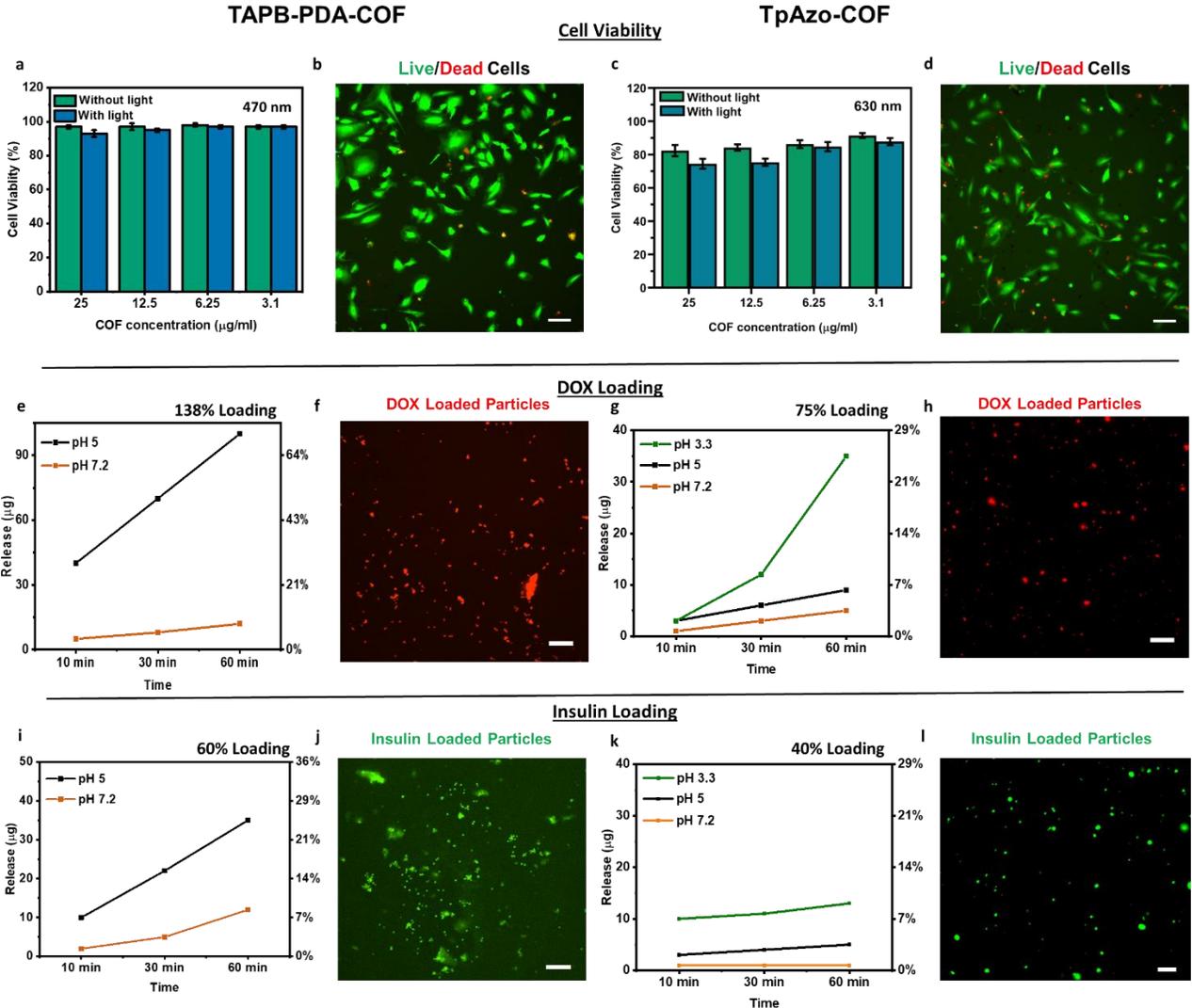

**Figure 3: COF microswimmer biocompatibility, drug loading, and triggered release properties. a-d:** In vitro cell viability results for COF microswimmers **a, c:** cell viability percentages of HUVEC cells in the presence of increasing TAPB-PDA-COF and TpAzo-COF microswimmer concentrations with/without 470 nm and 630 nm illumination, respectively, for 30 minutes, mean ± S.D. **b, d:** Corresponding fluorescence images of live cells (green) and dead cells (red) with 25 μg/ml, 30 minutes, 470 nm and 630 nm, respectively. **e-h:** DOX uptake & release results for COF microswimmers. **e:** TAPB-PDA-COF loading and release capacity with Doxorubicin (DOX) in MEM at different pH over time, reaching 138% for TABP-PDA-COF loaded in MEM. **f:** Corresponding fluorescence image of DOX (red) loaded TAPB-PDA-COFs at 25 μg/ml concentration. **g:** TpAzo-COF with 75% loading and their subsequent stepwise release at different pH conditions; in neutral pH (7.2), slightly acidic conditions (pH=5), and acidic (3.3) as encountered around cancer cells. **h:** Corresponding fluorescence image of DOX (red) loaded TpAzo-COFs at 25 µg/ml concentration. **i-l:** Insulin uptake & release results for COF microswimmers. **i:** Insulin loading of TAPB-PDA-COF with 60 % loading in MEM and release in different pH values over time. **j:** Corresponding fluorescence images of FITC (green) labeled insulin-loaded TAPB-PDA COFs. **k:** Insulin loading of TpAzo-COF with 40% loading in MEM and its release at different pH values over time. **l:** Corresponding fluorescence images of FITC (green) labeled insulin-loaded TpAzo-COFs. All scale bars are 100 µm.

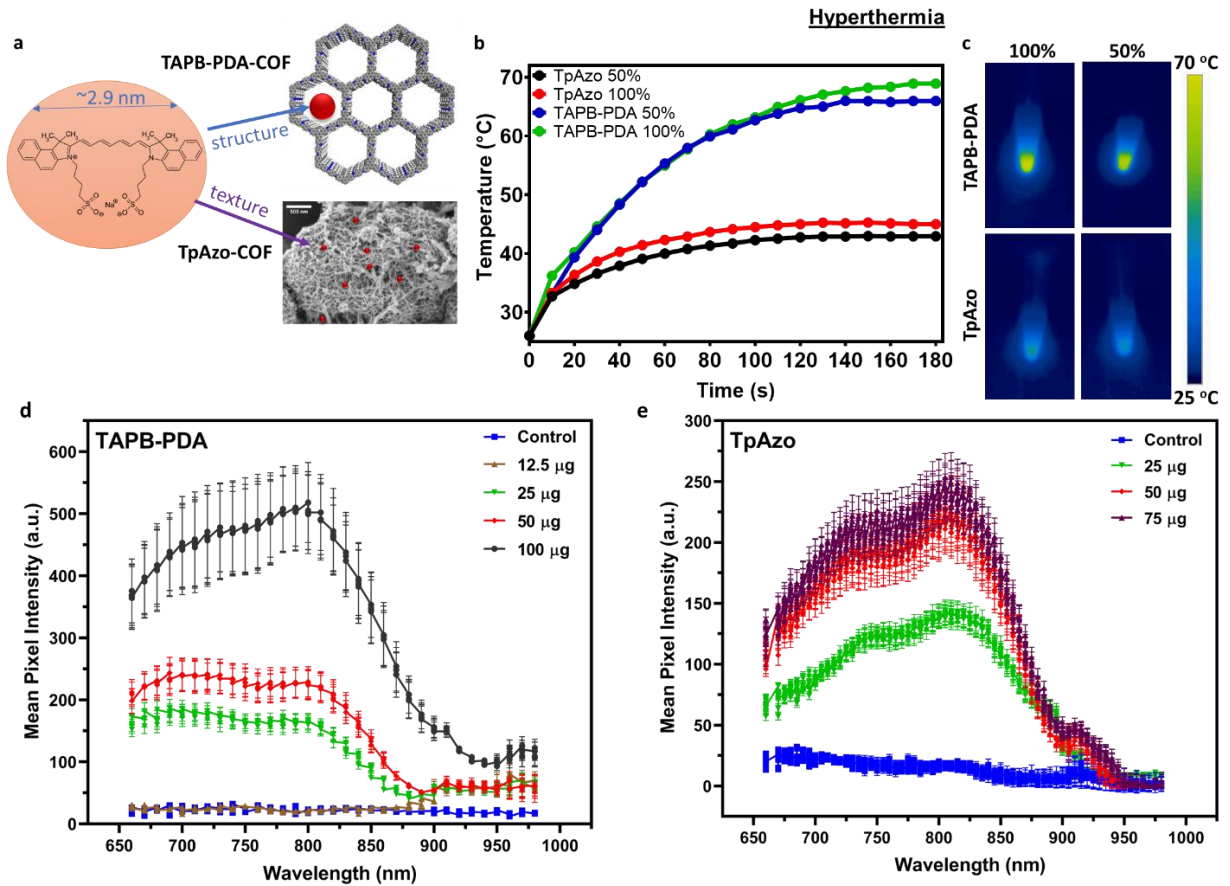

**Figure 4: Indocyanine green (ICG) loading, imaging, and hyperthermia functions of both COF microswimmer types. a:** ICG uptake into suitable structural pores (TAPB-PDA-COF) or texturally porous structures (Tp-Azo-COF). **b,c**: NIR-based heating of 50% and 100% ICG-loaded COF particles. **d**: Intensity of photoacoustic signal vs. ICG loading, highlighting high sensitivity regimes at low loading concentrations for TAPB-PDA-COF microswimmers. **e**: The photoacoustic signal intensity vs. ICG loading highlights high sensitivity regimes at low loading concentrations for TpAzo COF microswimmers.

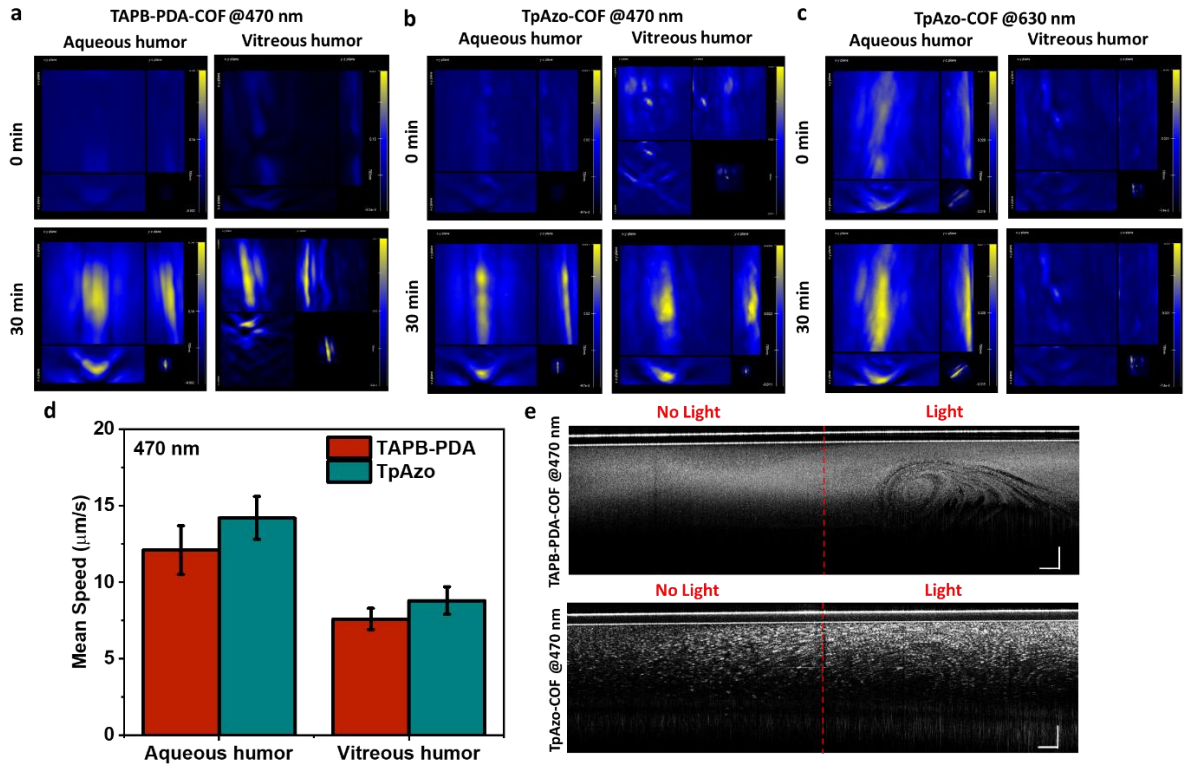

**Figure 5: Real-time imaging of COF motion by photoacoustic and optical coherence tomography imaging modalities. a-c:** Photoacoustic imaging of focused light-driven actuation of ICG-loaded COFs in both intraocular fluids. After 30 min, the accumulation of COF microswimmers in the focus of the light with different wavelengths is visible. **d:** Mean speeds of COF microswimmer particles illuminated with 470 nm light in intraocular fluids (**Video S3**). **e:** Optical coherence images of COFs in aqueous humor (**Video S4**). The COF swimmers' light-driven movement on the tubing's light-applied side is visible. The scale bar is 500 µm on each axis.

Supporting Information

# Designing Covalent Organic Framework-based Light-driven Microswimmers towards Intraocular Theranostic Applications

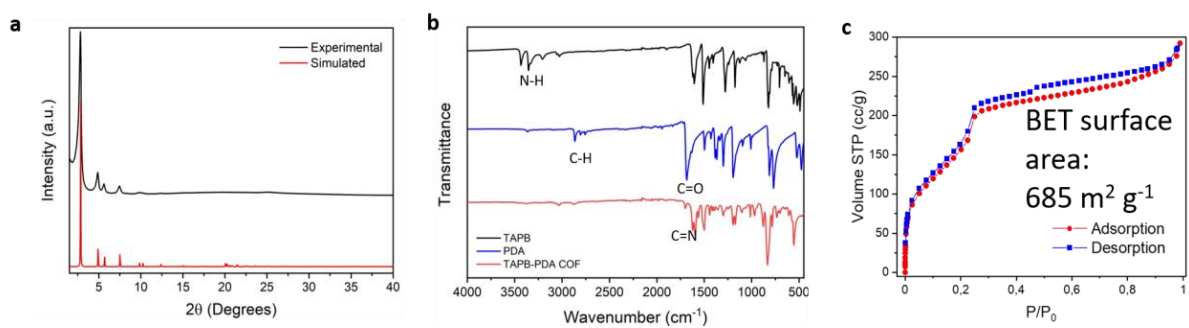

**Figure S1**. TABP-PDA COF structural analysis. **a:** Powder XRD after washing. **b:** FT-IR of the precursors and the COF. **c:** BET surface area measurement for overall surface area analysis.

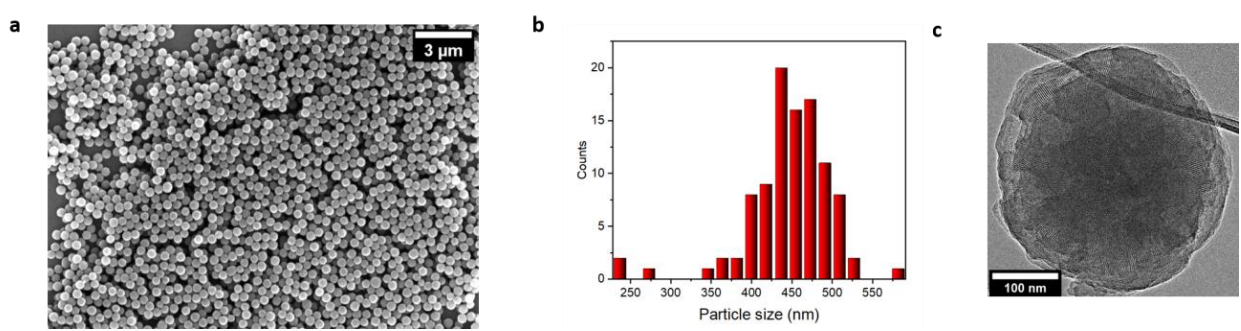

**Figure S2**. TABP-PDA COF particle morphology and structure. **a:** SEM image illustrating uniform size distribution of the washed COF microparticles. **b:** Particle size distribution showing high uniformity. **c:** TEM image showing a single COF nanoparticle consisting of crystalline domains with a lateral size of approx. 50 nm.

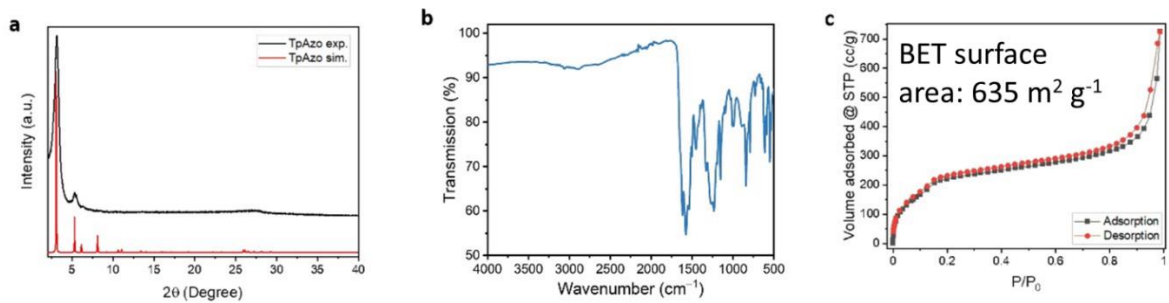

**Figure S3**: TpAzo-COF structural analysis. **a:** Powder XRD after washing. **b**: FTIR of the COF. **c:** BET surface area measurement for overall surface area analysis.

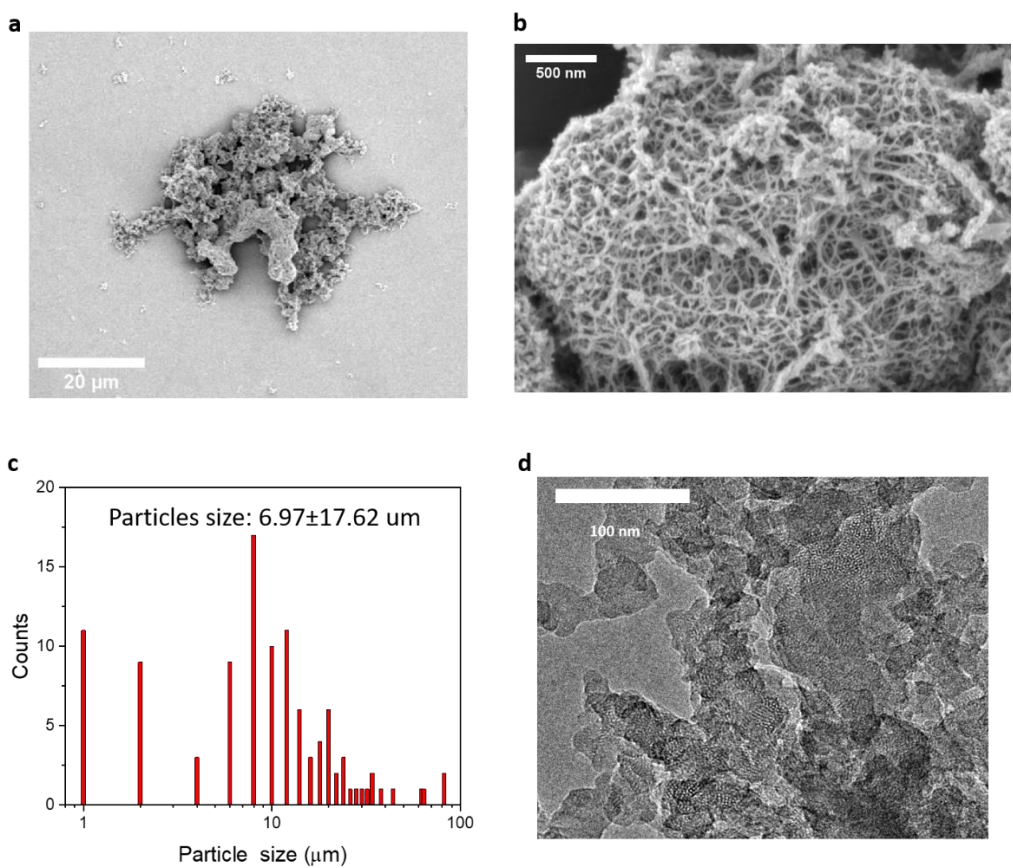

**Figure S4**. TpAzo-COF particle morphology and structure. **a:** SEM image illustrating the agglomerated structure of TpAzo-COF microparticles. **b:** SEM image (zoomed in) showing sponge-like inner structure with macropores. **c:** Particle size distribution showing non-uniformity of the particle agglomerates. The particle size is centered around 7 μm. **d:** TEM image showing the interconnection of crystalline COF nanosheets with a domain size of approximately 50 nm or less.

## Supporting Videos

**Video S1.** Light-driven propulsion of 100 µg/ml TABP-PDA and TpAzo COF microswimmers inside distilled water with a 470-nm wavelength light source

**Video S2.** Phototaxis behavior of TABP-PDA and TpAzo COF microswimmers inside MEM using a directional 470-nm wavelength light source

**Video S3.** TABP-PDA COF and TpAzo COF microswimmer propulsion inside the porcine aqueous and porcine vitreous humor fluid

**Video S4.** Optical coherence tomography (OCT) imaging and guided trapping of TABP-PDA and TpAzo COF microswimmers inside the aqueous humor fluid

**Video S5.** Optical coherence tomography (OCT) imaging and guided propulsion of TABP-PDA and TpAzo COF microswimmers inside the anterior chambers of the porcine eye